\documentclass[aps,nofootinbib,showpacs,twocolumn,floatfix]{revtex4}
\usepackage{epsf,amsfonts,amssymb,amsbsy}
\newcommand{\Rsub}{\rm\scriptscriptstyle}
\begin{document}
\title{Leptonic constant of pseudoscalar $\boldsymbol B_{\boldsymbol c}$ meson}
\author{V.V.Kiselev}
\email{kiselev@th1.ihep.su}
\affiliation{Russian State Research Center "Institute for High
Energy
Physics",
Protvino, Moscow Region, 142280, Russia\\ Fax: +7-0967-744739}
\begin{abstract}
We calculate the leptonic constant for the ground pseudoscalar
state of $B_c$ meson in the framework of QCD-motivated potential
model taking into account the two-loop anomalous dimension for the
heavy quark current in the nonrelativistic QCD as matched with the
full QCD.
\end{abstract}
\maketitle

\hbadness=1500
\section{Introduction}
The investigation of long-lived heavy quarkonium $B_c$ composed of
quarks with different flavors can produce a significant progress
in the study of heavy quark dynamics, since the variation of bound
state conditions for the heavy quarks in various systems such as
the heavy-light hadrons or doubly heavy mesons and
baryons\footnote{See the review of physics with the baryons
containing two heavy quarks in \cite{QQq}.} provides us with the
different conditions in both the binding of quarks by the strong
interactions and the electroweak decays. In addition to the rich
fields of study such as the spectroscopy, production mechanism and
lifetime there is a possibility to get the model-independent
information on the CP-violating parameters in the heavy quark
sector\footnote{The extraction of angle $\gamma$ in the unitary
triangle derived from the charged current mixing in the heavy
quark sector can be obtained from the decays of doubly heavy
baryons, too, in the same manner \cite{CPQQq}.} \cite{CPBc}.

The first experimental observation of the $B_c$ meson by the CDF
collaboration \cite{cdf} confirmed the theoretical predictions on
its mass, production rate and lifetime \cite{revbc}. So, we could
expect, that an essential increase of statistics in the nearest
future will provide us with a new battle field in the study of
long-lived doubly heavy hadrons at Tevatron \cite{BRunII} and LHC
\cite{LHCB}.

The special role of heavy quarks in QCD is caused by the small
ratio of two physical scales: the heavy quark mass $m_Q$ is much
greater than the energetic scale of the quark confinement
$\Lambda_{\mbox{\sc qcd}}$. This fact opens a way to develop two
powerful tools in the description of strong interactions with the
heavy quarks. The first tool is the perturbative calculations of
Wilson coefficients determined by the hard corrections with the
use of renormalization group improvements. The second instrument
is the Operator Product Expansion (OPE) related with the small
virtuality of heavy quark in the bound state, which reveals itself
in many faces such as the general expansion of operators in
inverse powers of heavy quark mass and the QCD sum rules. A
specific form of OPE is the application for the heavy quark
lagrangian itself, which results in the effective theories of
heavy quarks. The effective theory is constructed under the choice
of its leading term appropriate for the system under study. So,
the kinetic energy can be neglected in the heavy-light hadrons to
the leading order. The corresponding effective theory is called
HQET \cite{HQET}. In the doubly heavy mesons the kinetic energy is
of the same order as the potential one, while the velocity of the
heavy quark motion is small, and we deal with the nonrelativistic
QCD \cite{NRQCD} and its developments by taking into account the
static energy in the potential NRQCD in a general form (pNRQCD)
\cite{pNRQCD} or under the correlation of quarkonium size and the
time for the formation of bound system with the improved scheme of
expansion in the heavy quark velocity (vNRQCD) \cite{vNRQCD}. The
perturbative QCD remains actual in the effective theories, since
it is necessary for the calculation of Wilson coefficients
determined by hard corrections. More definitely, the perturbative
calculations determine both the matching of Wilson coefficients in
the effective theory with the full QCD and the anomalous
dimensions resulting in the evolution of the coefficients with the
variation of normalization point. In this respect we have to
mention that the pNRQCD results on the static potential and the
mass-dependent terms in the heavy quark-antiquark energy were
confirmed by vNRQCD after appropriate limits and some
calculational corrections in vNRQCD. In addition, the pNRQCD is a
powerful tool in the studies of both the spectroscopy and the
heavy quarkonium decays \cite{pNRQCD2}.

In contrast to the Wilson coefficients, the hadronic matrix
elements of operators composed by the effective fields of
nonrelativistic heavy quarks cannot be evaluated in the
perturbative manner. In this paper we use the potential model for
such the calculations in the case of leptonic constant for the
mixed-flavor heavy quarkonium $B_c$. The matching procedure is
performed with the two-loop accuracy available to the moment.

\section{Leptonic constant of $\boldsymbol B_{\boldsymbol c}$ in the potential approach}

In the NRQCD approximation for the heavy quarks, the calculation of leptonic
constant for the heavy quarkonium with the two-loop accuracy requires the
matching of NRQCD currents with the currents in full QCD,
$$
J_\nu^{\Rsub QCD}= \bar Q_1 \gamma_5\gamma_\nu Q_2, \;\;\; {\cal
J}_\nu^{\Rsub NRQCD} = -\chi^\dagger  \phi \; v_\nu,
$$
where we have introduced the following notations: $Q_{1,2}$ are the
relativistic quark fields, $\chi$ and $\phi$ are the nonrelativistic spinors of
anti-quark and quark, $v$ is the four-velocity of heavy quarkonium, so that
\begin{equation}
J_\nu^{\Rsub QCD} = {\cal K}(\mu_{\rm hard}; \mu_{\rm fact})\cdot
{\cal J}_\nu^{\Rsub NRQCD}(\mu_{\rm fact}), \label{match}
\end{equation}
where the scale $\mu_{\rm hard}$ gives the normalization point for the matching
of NRQCD with full QCD, while $\mu_{\rm fact}$ denotes the normalization point
for the calculations in the perturbation theory of NRQCD.

For the pseudoscalar heavy quarkonium composed of heavy quarks
with the different flavors, the Wilson coefficient ${\cal K}$ is
calculated with the two-loop
accuracy  
\begin{eqnarray}
{\cal K}(\mu_{\rm hard}; \mu_{\rm fact}) &=& 1 + c_1\,
\frac{\alpha_s^{\overline{\Rsub MS}}(\mu_{\rm
hard})}{\pi}+\nonumber\\ && c_2(\mu_{\rm hard}; \mu_{\rm
fact})\left(\frac{\alpha_s^{\overline{\Rsub MS}}(\mu_{\rm
hard})}{\pi}\right)^2\hspace*{-4pt}, \label{kfact}
\end{eqnarray}
and $c_{1,2}$ are explicitly given in Refs. \cite{braflem} and
\cite{ov}, respectively. The anomalous dimension of factor ${\cal K}(\mu_{\rm
hard}; \mu_{\rm fact})$ in NRQCD is defined by
\begin{equation}
\frac{d \ln{\cal K}(\mu_{\rm hard}; \mu)}{d \ln \mu} = \sum_{k=1}^{\infty}
\gamma_{[k]} \left(\frac{\alpha_s^{\overline{\Rsub MS}}(\mu)}{4\pi}\right)^k,
\label{anom}
\end{equation}
whereas the two-loop calculations\footnote{We use ordinary notations for the
invariants of $SU(N_c)$ representations: $C_F=\frac{N_c^2-1}{2 N_c}$, $C_A=
N_c$, $T_F = \frac{1}{2}$, $n_f$ is a number of ``active'' light quark
flavors.} give
\begin{eqnarray}
\gamma_{[1]} & = & 0,\\
\gamma_{[2]} & = & -8 \pi^2 C_F
\left[\left(2-\frac{(1-r)^2}{(1+r)^2}\right) C_F + C_A\right],
\label{g2}
\end{eqnarray}
where $r$ denotes the ratio of heavy quark masses. The initial
condition for the evolution of factor ${\cal K}(\mu_{\rm hard};
\mu_{\rm fact})$ is given by the matching of NRQCD current with
full QCD at $\mu_{\rm fact} = \mu_{\rm hard}$.

The leptonic constant is defined in the following way:
\begin{equation}
\langle 0| J_\nu^{\Rsub QCD} |\bar Q Q \rangle = v_\nu f_{\bar Q Q
} M_{\bar Q Q }.
\end{equation}
In full QCD the axial vector current of quarks has zero anomalous
dimension, while in NRQCD the current ${\cal J}_\nu^{\Rsub NRQCD}$
has the nonzero anomalous dimension, so that in accordance with
(\ref{match})--(\ref{g2}), we find
\begin{equation}
\langle 0| {\cal J}_\nu^{\Rsub NRQCD}(\mu) |\bar Q Q \rangle =
{\cal A}(\mu)\; v_\nu f_{\bar Q Q }^{\Rsub NRQCD} M_{\bar Q Q },
\label{a}
\end{equation}
where, in terms of nonrelativistic quarks, the leptonic constant for the heavy
quarkonium is given by the well-known relation with the wave function at the
origin
\begin{equation}
f_{\bar Q Q}^{\Rsub NRQCD} = \sqrt{\frac{12}{M_{\bar Q Q}}}\;
|\Psi_{\bar Q Q}(0)|, \label{wave}
\end{equation}
and the value of wave function in the leading order is determined
by the solution of Schr\"odinger equation with the static
potential, so that we isolate the scale dependence of NRQCD
current in the factor ${\cal A(\mu)}$, while the leptonic constant
$f_{\bar Q Q}^{\Rsub NRQCD}$ is evaluated at a fixed normalization
point $\mu=\mu_0$, which will be attributed below. It is evident
that
\begin{equation}
f_{\bar QQ} = f_{\bar QQ}^{\Rsub NRQCD} {\cal A}(\mu_{\rm
fact})\cdot {\cal K}(\mu_{\rm hard}; \mu_{\rm fact}), \label{cc}
\end{equation}
and the anomalous dimension of ${\cal A}(\mu_{\rm fact})$ should compensate the
anomalous dimension of factor ${\cal K}(\mu_{\rm hard}; \mu_{\rm fact})$, so
that in two loops we have got
\begin{equation}
\frac{d \ln{\cal A}(\mu)}{d \ln \mu} = - \gamma_{[2]}
\left(\frac{\alpha_s^{\overline{\Rsub MS}}(\mu)}{4\pi}\right)^2.
\label{anoma}
\end{equation}
The physical meaning of ${\cal A}(\mu)$ is clearly determined by the relations
of (\ref{a}) and (\ref{cc}): this factor gives the normalization of matrix
element for the current of nonrelativistic quarks expressed in terms of wave
function for the two-particle quark state (in the leading order of inverse
heavy quark mass in NRQCD). Certainly, in this approach the current of
nonrelativistic quarks is factorized from the quark-gluon sea, which is a
necessary attribute of hadronic state, so that, in general, this physical state
can be only approximately represented as the two-quark bound state. In the
consideration of leptonic constants in the framework of NRQCD, this
approximation requires to introduce the normalization factor ${\cal A}(\mu)$
depending on the scale.

The renormalization group equation of (\ref{anoma}) is simply integrated out,
so that
\begin{equation}
{\cal A}(\mu) = {\cal A}(\mu_0)\; \left[ \frac{\beta_0+\beta_1
{\displaystyle\frac{\alpha_s^{\overline{\Rsub
MS}}(\mu)}{4\pi}}}{\beta_0+\beta_1 {\displaystyle
\frac{\alpha_s^{\overline{\Rsub MS}}(\mu_0)}{4\pi}}}
\right]^{\displaystyle\frac{\gamma_{[2]}}{2\beta_1}}, \label{RG2}
\end{equation}
where $\beta_0 = \frac{11}{3} C_A - \frac{4}{3} T_F n_f$, and
$\beta_1 = \frac{34}{3} C_A^2 - 4 C_F T_F n_f  - \frac{20}{3} C_A
T_F n_f$. A constant of integration could be defined so that at a
scale $\mu_0$ we would get ${\cal A}(\mu_0)=1$. Thus, in the
framework of NRQCD we have got the parametric dependence of
leptonic constant estimates on the scale $\mu_0$, which has the
following simple interpretation: the normalization of matrix
element for the current of nonrelativistic quarks at $\mu_0$ is
completely given by the wave function of two-quark bound state. At
other $\mu\ne \mu_0$ we have to introduce the factor ${\cal
A}(\mu)\ne 1$, so that the approximation of hadronic state by the
two-quark wavefunction becomes inexact.

Following the method described in \cite{KKO,KLPS}, we estimate the wave
function of $\bar b c$ quarkonium in the nonrelativistic model with the static
potential given by \cite{KKO}, so that
\begin{equation}
\sqrt{4\pi}|\Psi(0)_{\bar b c}| = 1.26\;{\rm GeV}^{3/2}.
\end{equation}
In a wide region of distances between the quarks $\rho$, the static potential
of \cite{KKO} free of infrared singularity can be matched with the two-loop
perturbative potential $V_{\rm pert}$ calculated in \cite{Peter,Schroed}. In
this way the difference between the potentials
\begin{equation}
\delta V(\mu_{\rm soft}) = V_{\Rsub KKO}(\rho) - V_{\rm
pert}(\rho; \mu_{\rm soft})
\end{equation}
depends on the normalization point $\mu_{\rm soft}$ in the perturbative QCD of
static sources. It can be numerically approximated by the expression
$$
\frac{\delta V(\mu_{\rm soft})}{1\;{\rm GeV}} = \frac{0.99}{\mu_{\rm
soft}/(1\;{\rm GeV}) - 0.5},
$$
reflecting the infrared uncertainties in the perturbative calculations.

\begin{figure}[th]
\setlength{\unitlength}{0.6mm}
\begin{center}
\begin{picture}(100,110)
\put(-5,5){\epsfxsize=100\unitlength \epsfbox{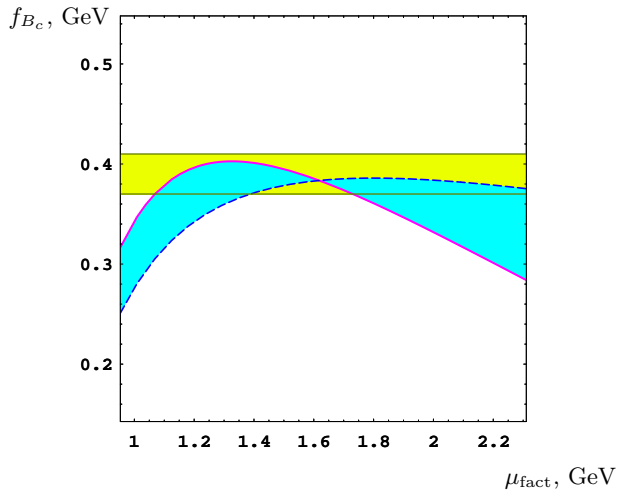}}
\put(90,0){$\mu_{\rm fact}$, GeV} \put(-20,102){$f_{B_c}$, GeV}
\end{picture}
\end{center}

\caption{The leptonic constant of ground pseudoscalar state in the
system of heavy quarkonium $\bar b c$ is presented versus the soft
scale of normalization. The shaded region restricted by curves
corresponds to the change of hard scale from $\mu_{\rm hard} = 3$
GeV (the dashed curve) to $\mu_{\rm hard} = 2$ GeV (the solid
curve) with the initial condition for the evolution of
normalization factor ${\cal A}(\mu_{\rm fact})$ posed in the form
of ${\cal A}(1.2\;{\rm GeV})=1$ and ${\cal A}(1.\;{\rm GeV})=1$,
respectively, in the matrix element of current given in the
nonrelativistic representation. The horizontal band is the limits
expected from the QCD sum rules \cite{sr} and scaling relations
for the leptonic constants of heavy quarkonia \cite{sc}. In the
cross-point, the leptonic constant of $B_c$ weakly depends on the
parameters given by the hard scale of matching as well as the
scale of the initial normalization.} \label{fB_c}
\end{figure}

The masses of heavy quarks used in the potential model of \cite{KKO}
\begin{equation}
m_c^{\Rsub V} = 1.468\;{\rm GeV,}\quad
m_b^{\Rsub V} = 4.873\;{\rm GeV,}
\label{mcmb}
\end{equation}
are consistent with the values of running masses as was shown in
\cite{KKO}. The cancellation of renormalon in the sum of the
perturbative static potential and pole masses of heavy quarks
takes place, so that\footnote{The accuracy of expression below,
i.e. a possible additive shift, is discussed in \cite{RGI}.}
$$
m_{\rm pole}(\mu_{\rm soft}) = m^{\Rsub V} + \frac{1}{2}\,\delta V(\mu_{\rm
soft}).
$$
In what follows we put the normalization point of perturbative potential
$\mu_{\rm soft} = \mu_{\rm fact}$.

The result of calculation for the leptonic constant of $B_c$ in
the potential approach is shown in Fig. \ref{fB_c}. We chose the
values of $\mu_{\rm hard}$ in a range, so that the stable value of
leptonic constant would be posed at the scale $\mu_{\rm fact}<
m_c^{\rm pole}$, which is the condition of consistency for the
NRQCD approach. Then one can observe the position of $\mu_{\rm
fact}\approx 1.5-1.6$ GeV, where the estimate of $f_{B_c}$ is
independent of the choice of $\mu_{\rm hard}$ and $\mu_0$, and the
maximal values at the different choices of $\mu_{\rm hard}$ are
also close to each other.

\begin{figure}[th]
\setlength{\unitlength}{0.75mm}
\begin{center}
\begin{picture}(100,85)
\put(-5,5){\epsfxsize=100\unitlength \epsfbox{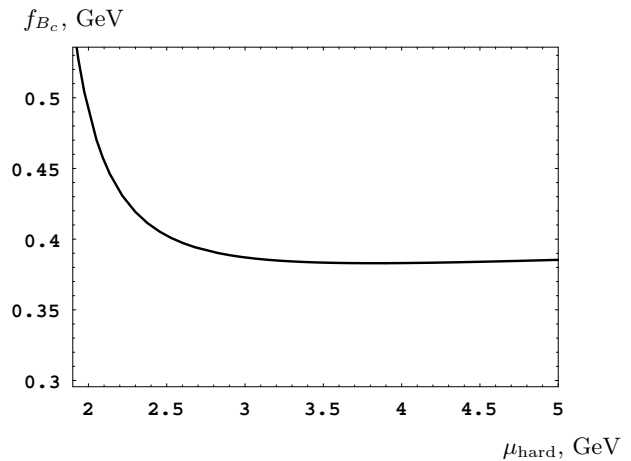}}
\put(83,1){$\mu_{\rm hard}$, GeV} \put(-2,77){$f_{B_c}$, GeV}
\end{picture}
\end{center}

\caption{The leptonic constant of ground pseudoscalar state in the
system of heavy quarkonium $\bar b c$ is presented versus the hard
scale of normalization at $\mu_{\rm fact} = 1.5$ GeV.}
\label{fB_chard}
\end{figure}

We have found that in order to reach the minimal dependence of leptonic
constant on the hard scale, the normalization point $\mu_0$ should correlate
with $\mu_{\rm hard}$ at fixed $\mu_{\rm fact}$. For illustration we show the
result at $\mu_{\rm fact} = 1.5$ GeV and
$$
\mu_0 = 1.6\;{\rm GeV} - \delta V(\mu_{\rm hard})
$$
in Fig. \ref{fB_chard}.

\begin{figure}[th]
\setlength{\unitlength}{0.7mm}
\begin{center}
\begin{picture}(115,95)
\put(-5,8){\epsfxsize=120\unitlength \epsfbox{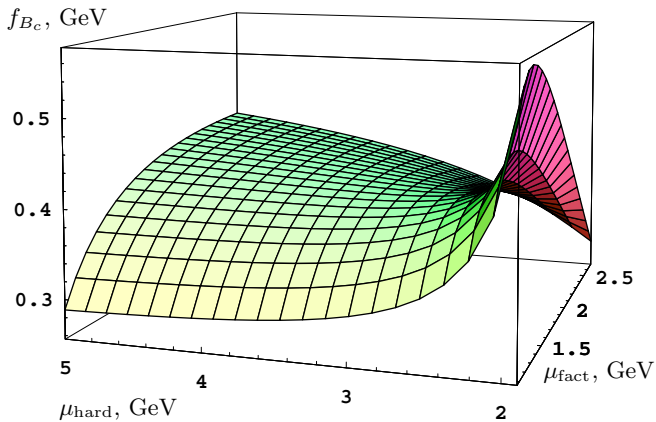}}
\put(97,20){$\mu_{\rm fact}$, GeV} \put(5,13){$\mu_{\rm hard}$,
GeV} \put(-5,87){$f_{B_c}$, GeV}
\end{picture}
\end{center}

\vspace*{-12mm}
\caption{The leptonic constant of ground pseudoscalar state in the system of
heavy quarkonium $\bar b c$ is presented versus the soft and hard scales,
$\mu_{\rm fact}$ and $\mu_{\rm hard}$. The initial condition for the evolution
of normalization factor ${\cal A}(\mu_{\rm fact})$ in the matrix element of
current in the nonrelativistic representation is given by ${\cal A}(1.6\;{\rm
GeV}-\delta V(\mu_{\rm hard}))=1$.}
\label{fB_c3d}
\end{figure}

The full dependence on the scales is presented in Fig.
\ref{fB_c3d}. We can see that there is a saddle stable point for
the leptonic constant depending on two scales: the normalization
point of perturbative calculation for the Wilson coefficient in
NRQCD and the matching point of NRQCD with the full QCD. This
stability implies that in the vicinity of saddle point the
contribution of higher corrections in $\alpha_s$ is not
significant, while at other values of scales these corrections
should be taken into account.

The final result of two-loop calculations is
\begin{equation}
f_{B_c} = 395\pm 15\; {\rm MeV}. \label{fin}
\end{equation}
It should be compared with the estimate of potential model itself
without the matching
\begin{equation}
f^{\Rsub NRQCD}_{B_c} = 493\; {\rm MeV},
\end{equation}
which indicates the magnitude of the correction about 20\%.
Furthermore, the calculations in the same potential model with the
one-loop matching \cite{KKO} gave
\begin{equation}
f^{1-loop}_{B_c} = 400\pm 45\; {\rm MeV},
\end{equation}
where the uncertainty is significantly greater than in the
two-loop procedure, since at the one-loop level we have no stable
point in the scale dependence of the result. Therefore, in
contrast to the discussion given in \cite{ov} we see that the
correction is not crucially large, but it is under control in the
system of $B_c$. The reason for such the claim on the reliability
of result is caused by two circumstances. First, the one-loop
anomalous dimension of NRQCD current is equal to zero. Therefore,
we start the summation of large logs in the framework of
renormalization group (RG) with the expressions in (\ref{anoma})
and (\ref{RG2}). Second, after such the summation of large logs
the three-loop corrections could be considered as small beyond the
leading RG logs.

The result on $f_{B_c}$ is in agreement with the scaling relation
derived from the quasi-local QCD sum rules \cite{sc}, which use
the regularity in the heavy quarkonium mass spectra, i.e. the fact
that the splitting between the quarkonium levels after the
averaging over the spins of heavy quarks weakly depends on the
quark flavors. So, the scaling law for the S-wave quarkonia has
the form
\begin{equation}
\frac{f_n^2}{M_n}\;\left(\frac{M_n}{M_1}\right)^2\;
\left(\frac{m_1+m_2}{4\mu_{12}}\right)^2
 = \frac{c}{n}, \label{2.3}
\end{equation}
where $n$ is the radial quantum number, $m_{1,2}$ are the masses
of heavy quarks composing the quarkonium, $\mu_{12}$ is the
reduced mass of quarks, and $c$ is a dimensional constant
independent of both the quark flavors and the level number $n$.
The value of $c$ is determined by the splitting between the $2S$
and $1S$ levels, or the average kinetic energy of heavy quarks,
which is independent of the quark flavors and $n$ with the
accuracy accepted. The accuracy depends on the heavy quark masses,
and it is discussed in \cite{sc} in detail. The parameter $c$ can
be extracted from the known leptonic constants of $\psi$ and
$\Upsilon$, so that the scaling relation gives
$$
f_{B^*_c}\approx 400\;{\rm MeV}
$$
for the vector state. The difference between the leptonic
constants for the pseudoscalar and vector $1S$-states is caused by
the spin-dependent corrections, which are small. Numerically, we
get $|f_{B^*_c}-f_{B_c}|/f_{B^*_c}< 3\%$, hence, the estimates
obtained from the potential model and the scaling relation is in a
good agreement with each other.

\section{Conclusion}

In this paper we have calculated the leptonic constant for the
ground state of $\bar b c$ system in the framework of the
QCD-motivated potential model used for the estimate of
nonperturbative hadronic matrix element for the effective fields
of nonrelativistic quarks. The Wilson coefficient calculated with
the two-loop accuracy \cite{ov} has been implemented to relate the
currents of heavy quarks in the full QCD and NRQCD. It depends on
two scales: the matching point and the normalization scale in
NRQCD. The stable point in the scale parameter space has been
observed, which makes the estimate to be reliable. We have found
that the two-loop corrections are under control in the $B_c$
meson.

The numerical value of $f_{B_c}$ is in agreement with the
estimates obtained in the framework of QCD sum rules \cite{sr},
potential models \cite{revbc} and from the scaling relation
\cite{sc}. The expected systematic error coming from the usage of
potential model is reduced due to the QCD motivation consistent
with the measurements of QCD coupling constant at large
virtualities, the asymptotic freedom up to three loops and the
fitting the mass spectra of charmonia and bottominia, so that the
systematic accuracy is of the same order as the uncertainty
related with the variation of parametric scales shown in
(\ref{fin}).

The author thanks A.Onishchenko for a possibility to get fresh
results on the two-loop Wilson coefficient in the language of
MATHEMATICA package and for valuable remarks. I am grateful to
Antonio Vairo for the discussions on the results of pNRQCD and
vNRQCD.

This work is in part supported by the Russian Foundation for Basic
Research, grants 01-02-99315, 01-02-16585.

\end{document}